\documentclass[10pt,twocolumn,letterpaper]{article}

\usepackage{cvpr}              % To produce the CAMERA-READY version
\usepackage{siunitx}
\usepackage{mathtools}
\usepackage{algorithm}
\usepackage{algpseudocode}
\usepackage[dvipsnames]{xcolor}

\definecolor{cvprblue}{rgb}{0.21,0.49,0.74}
\usepackage[pagebackref,breaklinks,colorlinks,citecolor=cvprblue]{hyperref}
\usepackage{amsthm}			% theorems
\usepackage{amssymb}			% varnothing
\usepackage{algorithm}
\usepackage{bm}
\newcolumntype{L}[1]{>{\raggedright\arraybackslash}p{#1}}
\newcolumntype{C}[1]{>{\centering\arraybackslash}p{#1}}
\newcolumntype{R}[1]{>{\raggedleft\arraybackslash}p{#1}}

\theoremstyle{plain} % plain = italic, definition = roman

\def\E{\mathbb{E}}

\def\xbm{{\bm{x}}}
\def\zbm{{\bm{z}}}

\def\zbm{{\bm{z}}}

\def\thetabm{{\bm{\theta }}}

\def\Ncal{{\mathcal{N}}}

\newcommand{\norm}[1]{\left\lVert#1\right\rVert}

\def\paperID{4533} % *** Enter the Paper ID here
\def\confName{CVPR}
\def\confYear{2024}

\title{DDPET-3D: Dose-aware Diffusion Model for 3D Ultra Low-dose PET Imaging}

\author{Huidong Xie$^*$\\
Yale University\\
{\tt\small huidong.xie@yale.edu}
\and
Weijie Gan$^*$\\
Washington University in St. Louis\\
{\tt\small weijie.gan@wustl.edu}
\and
Bo Zhou\\
Yale University\\
{\tt\small bo.zhou@yale.edu}
\and
Xiongchao Chen\\
Yale University\\
{\tt\small xiongchao.chen@yale.edu}
\and
Qiong Liu\\
Yale University\\
{\tt\small qiong.liu@yale.edu}
\and
Xueqi Guo\\
Yale University\\
{\tt\small xueqi.guo@yale.edu}
\and
Liang Guo\\
Yale University\\
{\tt\small liang.guo@yale.edu}
\and
Hongyu An\\
Washington University in St. Louis\\
{\tt\small hongyuan@wustl.edu}
\and
Ulugbek S. Kamilov\\
Washington University in St. Louis\\
{\tt\small kamilov@wustl.edu}
\and
Ge Wang\\
Rensselaer Polytechnic Institute\\
{\tt\small wangg6@rpi.edu}
\and
Chi Liu\\
Yale University\\
{\tt\small chi.liu@yale.edu}
}

\begin{document}
\maketitle
\let\thefootnote\relax\footnotetext{$^*$ Equal Contributions}
\begin{abstract}
As PET imaging is accompanied by substantial radiation exposure and cancer risk, reducing radiation dose in PET scans is an important topic. Recently, diffusion models have emerged as the new state-of-the-art generative model to generate high-quality samples and have demonstrated strong potential for various tasks in medical imaging. However, it is difficult to extend diffusion models for 3D image reconstructions due to the memory burden. Directly stacking 2D slices together to create 3D image volumes would results in severe inconsistencies between slices. Previous works tried to either apply a penalty term along the z-axis to remove inconsistencies or reconstruct the 3D image volumes with 2 pre-trained perpendicular 2D diffusion models. Nonetheless, these previous methods failed to produce satisfactory results in challenging cases for PET image denoising. In addition to administered dose, the noise levels in PET images are affected by several other factors in clinical settings, e.g. scan time, medical history, patient size, and weight, etc. Therefore, a method to simultaneously denoise PET images with different noise-levels is needed. Here, we proposed a Dose-aware Diffusion model for 3D low-dose PET imaging (DDPET-3D) to address these challenges. We extensively evaluated DDPET-3D on 100 patients with 6 different low-dose levels (a total of 600 testing studies), and demonstrated superior performance over previous diffusion models for 3D imaging problems as well as previous noise-aware medical image denoising models. The code is available at: \href{https://github.com/xxx/xxx}{https://github.com/xxx/xxx}.
\end{abstract}    
\section{Introduction}
\label{sec:intro}

Positron Emission Tomography (PET) is a functional imaging modality widely used in oncology, cardiology, and neurology studies \cite{rohren2004clinical,schwaiger2005pet,clark2012cerebral}. Given the growing concern about radiation exposure and cancer risks accompanied with PET scans, reducing the PET injection dose is desirable \cite{robbins2008radiation}. However, PET image quality is negatively affected by the reduced injection dose and may affect diagnostic performance such as the identification of low-contrast lesions \cite{schaefferkoetter2015initial}. Therefore, reconstructing high-quality images from noisy input is an important topic.

Iterative methods such as Maximum Likelihood Expectation Maximization (MLEM) \cite{shepp_maximum_1982} and Ordered Subset Expectation Maximization (OSEM) \cite{hudson_accelerated_1994} are commonly used for PET reconstructions. However, they are vulnerable to noise in low-dose PET data. Iterative methods incorporating a Total Variation (TV) regularization could be used for noise reduction \cite{wang_low_2014}. However, they are time-consuming and often fail to produce high-quality reconstructions.

Deep learning has emerged as a new reconstruction algorithm for medical imaging tasks \cite{wang_perspective_2016}, and many different deep-learning methods were proposed for low-dose PET image reconstructions \cite{xu_200x_2017, zhou2020supervised, ouyang_ultra-low-dose_2019, zhou2021mdpet, zhou2023fast, zhou2022federated, zhou2023fedftn, liu2022personalized}. Recently, diffusion models have become the new state-of-the-art generative models~\cite{Croitoru.etal2023,Kazerouni.etal2023a}. They are capable of generating high-quality samples from Gaussian noise input, and have demonstrated strong potential for low-dose PET imaging. 
 For example, using the DDPM framework \cite{ho2020denoising}, Gong \textit{et al.} proposed to perform PET image denoising with MRI as prior information for improved image quality \cite{gong_pet_2023}. However, PET-MR systems are not widely available and it is difficult to obtain paired PET-MR images in reality for network training and image reconstructions. Jiang \textit{et al.}\cite{jiang_pet-diffusion_2023} adopted the latent diffusion model \cite{rombach_high-resolution_2022} for unsupervised PET denoising. Moreover, diffusion models have also been proposed for other imaging modalities, such as CT \cite{gao_corediff_2023} and MRI \cite{gungor_adaptive_2023, chung_score-based_2022}.

 However, these previous diffusion papers focus on 2D and do not address the 3D imaging problem. This is particularly important for PET imaging as PET is intrinsically a 3D imaging modality. A 3D diffusion model is desirable, however, due to the hardware memory limit, directly extending the diffusion model to 3D would be difficult. There are a few previous works that aim to address the 3D imaging problems of diffusion models. For example, Chung \textit{et al.}\cite{chung_solving_2023} proposed to apply a TV penalty term along the z-axis to remove inconsistencies within each reverse sampling step in the diffusion model. On the other hand, Lee \textit{et al.}\cite{lee_improving_2023} utilizes 2 pre-trained perpendicular 2D diffusion models to remove inconsistencies between slices. Nonetheless, these previous methods failed to produce satisfactory results in challenging cases for 3D PET image denoising. 

 Another challenge with PET image denoising is the high variation of image noise. The final noise level in PET images can be affected by different factors: \emph{(1)} Variations of acquisition start time. \emph{(2)} Occasional tracer injection infiltration in some patients, causing the tracer stuck in the arm, resulting in high image noise in the body. \emph{(3)} The variation of injection dose based on the patient's medical history, weight, and size. \emph{(4)} Different hospitals have different standard scan times for patients, resulting in variability of noise levels in the reconstructed images. Because of these reasons, a method to simultaneously denoise images with varying noise levels is desirable. However, previously proposed methods mentioned above have limited generalizability to different noise levels \cite{xie_unified_2023}. A network trained on one noise level fails to produce high-quality reconstructions on other noise levels. To address this problem, Xie \textit{et al. } proposed to combine multiple U-net-based \cite{ronneberger_u-net_2015} sub-networks with varying denoising power to generate optimal results for any input noise levels \cite{xie_unified_2023}. However, training multiple sub-networks requires tedious data pre-processing and long training time. The testing time also linearly increases with the number of sub-networks. Also, paired training data with different low-count levels may not be readily available. 

Moreover, different from other imaging modalities, PET was developed as a quantitative tool for clinical diagnosis. PET quantitative characteristics are increasingly being recognized as providing an objective, and more accurate measure for prognosis and response monitoring purposes than visual inspection alone \cite{boellaard_standards_2009}. However, our experimental results showed that, although standard diffusion models (DDPM \cite{ho2020denoising}, DDIM\cite{song_denoising_2022}) produce visually appealing reconstructions, they typically failed to maintain accurate quantification. For example, the total activities change after diffusion models. This is particularly true for whole-body PET scans since the tracer uptake could vary significantly in different organs. This problem was not able to address by simple normalization of the images. 

% In this work, we propose a dose-aware diffusion model for 3D PET denoising (DDPET-3D) to address these limitations and to improve denoising performance. In short, the main contribution of this paper is designing a diffusion model that can \emph{(1)} achieve noise-aware denoising. The proposed method was tested on PET images with 6 different low-count levels, ranging from 1\% to 50\%. \emph{(2)} address inconsistencies between slices while maintaining similar memory burden to 2D diffusion models. So that the proposed method does not require a large GPU cluster to train. \emph{(3)} produce denoised images that are not only visually appealing but also quantitatively accurate. \emph{(4)} fast-sampling. The proposed DDPET-3D can converge within 25 sampling steps. It reconstructs the entire 3D image volume with roughly 15 minutes on one NVIDIA A40 GPU (the matrix size is $673\times360\times360$). Standard DDPM sampling takes roughly 6 hours under the same settings. \emph{(5)} the proposed method was extensively tested on 100 patient studies, each with 6 different low-count levels, resulting in a total of 600 studies for evaluation.

In this work, we developed a dose-aware diffusion model for 3D PET Imaging (DDPET-3D) to address these limitations. The main contributions of the proposed DDPET-3D framework are: \emph{(1)} We proposed a dose-embedding strategy that allows noise-aware denoising. DDPET-3D can simultaneously denoise 3D PET images with varying low-dose/count levels. \emph{(2)} We proposed a 2.5D diffusion strategy with multiple fixed noise variables to address the 3D inconsistency issue between slices. DDPET-3D maintains a similar memory burden to 2D diffusion models while achieving high-quality reconstructions. \emph{(3)} We proposed to use a denoised prior in DDPET-3D, allowing it to converge within 25 sampling steps. DDPET-3D can denoise 3D PET images within a reasonable time constraint ($15$ mins on a single GPU). Previous diffusion methods using DDPM sampling would require approximately $6$ hrs. Evaluated on real-world ultra-low-dose 3D PET data, DDPET-3D demonstrates superior quantitative and qualitative results as compared with previous baseline methods. External validation of data from another country also shows our trained model can be reasonably generalized to other sites.

\section{Methods}
\label{sec:formatting}

% All text must be in a two-column format.
% The total allowable size of the text area is $6\frac78$ inches (17.46 cm) wide by $8\frac78$ inches (22.54 cm) high.
% Columns are to be $3\frac14$ inches (8.25 cm) wide, with a $\frac{5}{16}$ inch (0.8 cm) space between them.
% The main title (on the first page) should begin 1 inch (2.54 cm) from the top edge of the page.
% The second and following pages should begin 1 inch (2.54 cm) from the top edge.
% On all pages, the bottom margin should be $1\frac{1}{8}$ inches (2.86 cm) from the bottom edge of the page for $8.5 \times 11$-inch paper;
% for A4 paper, approximately $1\frac{5}{8}$ inches (4.13 cm) from the bottom edge of the
% page.

%-------------------------------------------------------------------------
\subsection{Diffusion Models}
The general idea of diffusion models is to learn the target data distribution $q(\xbm_0)$ (\emph{i.e.,} full-dose PET images in our case) using neural network. Once the distribution is learned, we can synthesize a new sample from it. Diffusion models consist of two Markov chains: the forward diffusion process and the learned reverse diffusion process.
The forward diffusion process $q$ gradually adds small amount of Gaussian noise to $\xbm_0\sim q(\xbm_0)$ in each step, until the original image signal is completely destroyed. As defined in the DDPM paper \cite{ho2020denoising} 
\begin{equation}
\begin{split}
q(\xbm_{1:T}|\xbm_0)&\coloneqq \prod_{t=1}^Tq(\xbm_t|\xbm_{t-1})\ , \\ \text{where} \; q(\xbm_t|\xbm_{t-1})&\coloneqq\mathcal{N}(\xbm_t;\sqrt{1-\beta_t}\xbm_{t-1},\beta_t\mathbf{I})\ .
\label{eq1}
\end{split}
\end{equation}
One property of the diffusion process is that, one can sample $\xbm_t$ for any arbitrary time-step $t$ without gradually adding noise to $\xbm_0$. By denoting $\alpha_t \coloneqq 1-\beta_t$ and $\bar{\alpha}_t \coloneqq \prod_{s=1}^t \alpha_s$, we have
\begin{equation}
\begin{split}
q(\xbm_t|\xbm_0)&=\mathcal{N}(\xbm_t;\sqrt{(\bar{\alpha}_t})\xbm_0,(1-\bar{\alpha}_t)\mathbf{I})\ , \\ 
\text{and}\ \xbm_t & = \sqrt{\bar{\alpha}_t}\xbm_0+\sqrt{1-\bar{\alpha}_t}{\bm \epsilon}\ ,
\label{eq3}
\end{split}
\end{equation}
where ${\bm \epsilon} \sim \mathcal{N}(0,\mathbf{I})$. The latent $\xbm_T$ is nearly an isotropic Gaussian distribution for a properly designed $\beta_t$ schedule. Therefore, one can easily generate a new $\xbm_T$ and then synthesize a $\xbm_0$ by progressively sampling from the reverse posterior $q(\xbm_{t-1}|\xbm_t)$.
However, this reverse posterior is tractable only if $\xbm_0$ is known
\begin{equation}
\label{equ:q-posi}
    q(\xbm_{t-1}|\xbm_t, \xbm_0) = \Ncal\Big(\xbm_{t-1}; \mu_q(\xbm_t,\xbm_0), \frac{\beta_t(1-\bar{\alpha}_{t-1})}{1-\bar{\alpha}_t}\textbf{I}\Big)\ ,
\end{equation}
where
\begin{equation}
    \mu_q(\xbm_t,\xbm_0) = \frac{\sqrt{\alpha_t}(1-\bar{\alpha}_{t-1})\xbm_t + \sqrt{\bar{\alpha}_{t-1}}(1-\alpha_t)\xbm_0}{1-\bar{\alpha}_t}\ .
\end{equation}
Note that $q(\xbm_{t-1}|\xbm_t):=q(\xbm_{t-1}|\xbm_t, \xbm_0)$, where the extra conditioning term $\xbm_0$ is superfluous due to the Markov property.
DDPM thus proposes to learn a parameterized Gaussian transitions $p_\thetabm(\xbm_{t-1}|\xbm_t)$ to approximate the reverse diffusion posterior \eqref{equ:q-posi}
\begin{equation}
\label{equ:p-posi}
    p_\thetabm(\xbm_{t-1}|\xbm_t) = \Ncal\Big(\xbm_{t-1}; \mu_\thetabm(\xbm_t,t), \sigma^2_t\textbf{I}\Big)\ ,
\end{equation}
where
\begin{equation}
    \label{equ:p-mean}
    \mu_\thetabm(\xbm_t,t) = \frac{1}{\sqrt{\alpha_t}}\Big(\xbm_t-\frac{1-\alpha_t}{\sqrt{1-\bar{\alpha}_t}}\epsilon_\thetabm(\xbm_t, t)\Big)\ .
\end{equation}
Here, $\epsilon_\thetabm$ denotes a neural network. 
% The training of $\epsilon_\thetabm$ is performed by optimizing the variational lower bound of the negative log likelihood. 
Through some derivations detailed in the DDPM paper \cite{ho2020denoising}, the training objective of $\epsilon_\thetabm(\xbm_t, t)$ can be formulated as follow
\begin{equation}
    \label{equ:training}
    \E_{\xbm,{\bm\epsilon},t\sim[1,T]}\big[\norm{{\bm \epsilon} - {\epsilon}_\thetabm(\xbm_t, t)}^2\big]\ .
\end{equation}
It is worth noting that the original DDPM set $\sigma_t$ to a fixed constant value based on the $\beta_t$ schedule. Recent studies~\cite{dhariwal2021diffusion,nichol2021improved} have shown the improved performance by using the learned variance $\sigma_t^2\coloneqq\sigma_\thetabm^2(\xbm_t, t)$. We also adopted this approach. To be specific, we have $\sigma_\thetabm(\xbm_t, t)\coloneqq\exp(v\log \beta_t + (1-v)\log\tilde{\beta}_t)$, where $\tilde{\beta}_t$ refers to the lower bounds for the reverse diffusion posterior variances \cite{ho2020denoising}, and $v$ denotes the network output. We used a single neural network with two seperate output channels to estimate the mean and the variance of \eqref{equ:p-mean} jointly.
Based on the learned reverse posterior $p_\thetabm(\xbm_{t-1}|\xbm_t)$, the iteration of obtaining a $\xbm_0$ from a $\xbm_T$ can be formulated as follow
\begin{equation}
    \xbm_{t-1} = \mu_\thetabm(\xbm_t,t) + \sigma_t\zbm,\text{ where } \zbm\sim\Ncal(0,\textbf{I})\ .
\end{equation}

\begin{figure*}[!ht]
\centerline{\includegraphics[width=\textwidth]{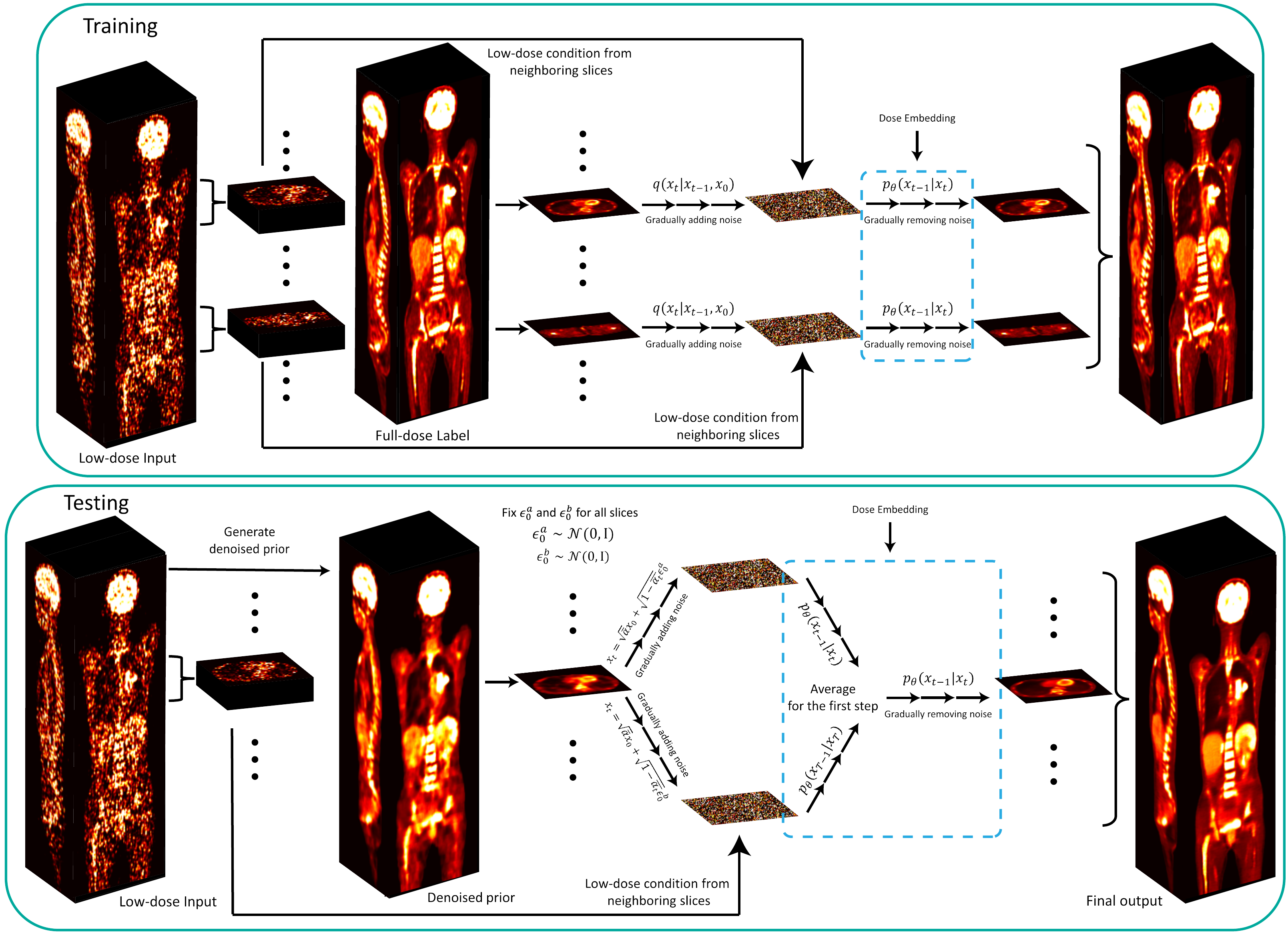}}
\caption{The training (\emph{top}) and the sampling (\emph{bottom}) pipeline of the proposed DDPET-3D method. The DDPET-3D uses multiple neighboring slices as additional inputs to predict the the central slice, allowing network to observe 3D information during training and testing. The DDPET-3D is also conditioned on the injected dose in order to accommodate inputs with varying noise levels. In sampling, we propose to fix the Gaussian latent for each slice in the 3D volume to address the inconsistency along z-axis. DDPET-3D also use a pre-trained denoiser prior to ensure accurate quantification in sampling.}
\label{fig_network}
\end{figure*}

\subsection{Conditional PET Image Denoising}

The framework described above only allows unconditional sampling. For the purpose of low-dose PET denoising, instead of generating new samples, the network needs to denoise the images based on input noisy counterparts. This can be achieved by adding additional condition to the neural network. $\epsilon_\thetabm{(\xbm_t,t)}$ becomes $\epsilon_\thetabm{(\xbm_t,t,\xbm_\mathrm{noisy})}$, where $\xbm_\mathrm{noisy}$ denotes the noisy input PET images. Specifically, the input to the neural network becomes a 2-channel input, one is $\xbm_t$, the other one is $\xbm_\mathrm{noisy}$. However, we noticed that, this technique results in severe inconsistencies in the reconstructed 3D image volumes. One example is presented in Fig.~\ref{fig_results_2view} (DDIM50+2.5D results).

\subsection{Proposed DDPET-3D}

The proposed framework is depicted in Fig.~\ref{fig_network}. The proposed network can observe 3D information from adjacent slices during the training process. This is achieved by using the $n$ neighboring slices as input to predict the central slice. Ablation studies showed that the performance converges at $n=31$, which was used in this work. One may use 3D convolutional layers to replace the 2D convolutional layers in the diffusion model for an enlarged receptive field and to allow the network to observe 3D structural information. However, we found that it significantly increases memory burden and makes the network difficult to optimize. To alleviate the memory burden and allow faster convergence, we embed neighboring slices in the channel dimension. Specifically, when using only one 2D slice as conditional information ($n=1$), the input dimension is $N_b\times W\times W\times2$, where the last dimension is the channel dimension, $N_b$ is input batch size, and $W$ is the width of the images. One channel is conditional 2D slice, and the other one is $\xbm_{t-1}$. When $n=31$, the input dimension becomes $N_b\times W\times W\times32$. 

Such technique allows the network to observe neighboring slices for 3D image reconstruction with only incremental increase in memory burden. Also, ablation studies showed that such technique also noticeably improves inconsistencies problems for 3D imaging. This is consistent with our expectation. With $n=1$, when the network tries to predict the next slice, we observe inconsistent reconstructions because both the conditional information and the starting Gaussian noise $p(\xbm_T)$ change. When $n=31$, the starting Gaussian noise $p(\xbm_T)$ still changes, but all the 31-channel conditional information only shifts a little. The network should produce more consistent output with only subtle changes in the input.

Another reason for the inconsistency issue is because the starting Gaussian noise of the reverse process is different for different slices. The diffusion sampling strategy starts from a random location in the high-dimensional space (random Gaussian noise), and approximates the data distribution $q(\xbm_0)$ based on the trained neural network, or the score function as described in \cite{song_score-based_2021}, through many iterations. The denoising problem is ill-posed, and we could generate an infinite number of different denoised images given the same low-count input with different starting Gaussian noise. This is beneficial for generative models to produce a wide variety of different images. However, the stochastic nature of the diffusion model would be problematic for medical image reconstruction problems because we expect neighboring slices to be consistent with each other in the 3D image volume. To address this problem, we proposed to fix the starting Gaussian noise when reconstructing all the slices in the entire 3D volume. Specifically, the starting Gaussian noise at the last time step $T$, $\xbm_T$ is fixed for all the slices during the sampling process, only the conditional low-count PET images change for different slices.

This approach produced more consistent reconstructions along the z-axis. However, we noticed that by fixing $p(\xbm_T)$, noise-dependent artifacts would propagate to all the slices along the z-axis. One example is presented in the first sub-figure in Fig.~\ref{fig_results_no_fix_seed_one_channel} (single $\epsilon$ image). To address this issue, as presented in Fig.~\ref{fig_network}, we proposed to initialize 2 different noise variables ${\bm\epsilon}_0^a$ and ${\bm\epsilon}_0^b$, and we have 2 different $\xbm_{t-1}$ in the reverse process. Fortunately, having 2 different noise variables does not significantly increase sampling time since we only need to average them at the first reverse step. This technique effectively addresses this issue.

As mentioned in the introduction section, despite the visually appealing results, the diffusion model usually produced images with inaccurate quantification. We noticed that, even though U-net-based methods failed to produce satisfactory results, especially for extremely low-dose settings, they typically maintained overall image quantification much better than diffusion models. To take advantage of that, instead of starting from random Gaussian noise, we first generate a denoised prior using a pre-trained U-net-based network, and then add Gaussian noise to it based on Equation \ref{eq3}. This added-noise denoised prior were used as the starting point of the reverse process. We adapted the Unified Noise-aware Network (UNN) proposed by Xie \textit{et al.} \cite{xie_unified_2023} to generate the denoised priors. 

For injected dose embedding, the values of the administered dose were first converted to Becquerel (Bq). Both $\sin$ and $\cos$ functions were then used to encode the injected dose and add with the diffusion time-steps (i.e., $t+\sin(\mathrm{dose}) + \cos(\mathrm{dose})$, where $t$ is the diffusion time-step). Time-step $t$ was also encoded using $\sin$ and $\cos$ functions. The encoded values were then fed into 2 linear layers to generate the embedding with Sigmoid linear unit (SiLU) in between.

For diffusion sampling, DDIM sampling enables faster convergence. But we noticed that using DDIM alone would tend to produce over-smoothed images. To alleviate this issue, adapted from the DDIM paper \cite{song_denoising_2022}, we proposed to have an interpolated sampling between DDIM and DDPM. We inserted DDPM samplings every 5 steps in between DDIM samplings to prevent the images from becoming over-smoothed. Algorithm \ref{alg1} displays the complete training procedure of the proposed method. Algorithm \ref{alg2} displays the complete sampling procedure of the proposed method.

\begin{algorithm}[t]
\caption{Training}\label{alg1}
\begin{algorithmic}
\State \textbf{Repeat}
\State $\xbm_0\sim q(\xbm_0)$ \Comment{Sample single-slice full-dose data}
\State $\xbm_{\mathrm{noisy}}\sim q(\xbm_{\mathrm{noisy}})$ \Comment{Sample multi-slice low-dose data}
\State $\mathrm{dose}_{\xbm_\mathrm{noisy}}$ \Comment{Injected dose of $\xbm_\mathrm{noisy}$} 
\State $t\sim \mathrm{Uniform({1,...,T})}$ \Comment{Sample diffusion time-step}
\State ${\bm \epsilon} \sim \mathcal{N}(0,\mathbf{I})$
\State $\nabla_\theta||{\bm \epsilon}-\epsilon_\thetabm(\sqrt{\bar{\alpha}_t}\xbm_0+\sqrt{1-\bar{\alpha}_t}{\bm \epsilon},t,\xbm_\mathrm{noisy}, \mathrm{dose}_{\xbm_\mathrm{noise}})$ \\ \hfill \Comment{Take gradient descent step}
\State \textbf{Until} convergence
\end{algorithmic}
\end{algorithm}

\begin{algorithm}[t]
\caption{Testing}\label{alg2}
\begin{algorithmic}
\State $\xbm_{\mathrm{noisy}}\sim q(\xbm_{\mathrm{noisy}})$ \Comment{Get entire low-dose image volume}
\State $\mathrm{dose}_{\xbm_\mathrm{noisy}}$ \Comment{Injected dose of $\xbm_\mathrm{noisy}$} 
\State $\xbm_{\mathrm{prior}} = \mathrm{UNN}(\xbm_{\mathrm{noisy}})$ \Comment{Generate denoised prior using} \\ \hfill {a pre-trained network (UNN) \cite{xie_unified_2023}}
% \State $(T',...,1) \gets \mathrm{ReSapce}(T,...,1)$ \Comment{Re-space time-steps}
\State $\xbm_\mathrm{recon}\gets \mathrm{zeros}(\xbm_{\mathrm{noisy}})$ \Comment{Initialize $\xbm_\mathrm{recon}$}
% \State $S$ \Comment{Total number of slices in $\xbm_\mathrm{recon}$}
\State ${\bm \epsilon}_0^a\sim \mathcal{N}(0,\mathbf{I})$, ${\bm \epsilon}_0^b\sim \mathcal{N}(0,\mathbf{I})$ \Comment{Obtain 2 noise variables}
\\
\While{$s=1,...,S$} \Comment{Total number of slices}
\State $\xbm_{\mathrm{noisy}}[s]$ \Comment{Conditioned multi-slice low-dose input}
\State $\xbm_{\mathrm{prior}}[s]$ \Comment{Conditioned single-slice denoised prior}
\State $\xbm_{T'}^a = \sqrt{\bar{\alpha}_{T'}}\xbm_{\mathrm{prior}}[s]+\sqrt{1-\bar{\alpha}_{T'}}{\bm \epsilon}_{0}^a$ 
\State $\xbm_{T'}^b = \sqrt{\bar{\alpha}_{T'}}\xbm_{\mathrm{prior}}[s]+\sqrt{1-\bar{\alpha}_{T'}}{\bm \epsilon}_{0}^b$ 
\\
\While{$t=T',...,1$}
\State $z \sim \mathcal{N}(0,\mathbf{I})$
\State $\xbm_{t-1}^a=\mathrm{Sampler}(\xbm_{t}^a,\xbm_{\mathrm{noisy}}[s],\mathrm{dose}_{\xbm_\mathrm{noisy}})$

\If{$t==T'$}
\State $\xbm_{t-1}^b=\mathrm{Sampler}(\xbm_{t}^b,\xbm_{\mathrm{noisy}}[s],\mathrm{dose}_{\xbm_\mathrm{noisy}})$
\State $\xbm_{t-1}^a = (\xbm_{t-1}^a + \xbm_{t-1}^b)/2$
\EndIf
\EndWhile
\State $\xbm_\mathrm{recon}[s] = \xbm_{1}^a$
\\
\EndWhile
\State \textbf{Return }$\xbm_\mathrm{recon}$
\end{algorithmic}
\end{algorithm}

\subsection{Low-count PET Data}
The PET dataset used in this study was collected at XX. 320 subjects with \textsuperscript{18}F-FDG tracer were acquired using a United Imaging uExplorer PET/CT system. The reconstruction matrix size is $673 \times 360\times 360$ with a $2.89\times 1.67 \times 1.67 \si{mm}^3$ voxel size. We randomly selected 210 subjects for training, 10 subjects for validation, and 100 subjects for testing. We downsampled the PET list-mode data to 1\%, 2\%, 5\%, 10\%, 25\%, and 50\% to simulate low-dose settings. All the images were reconstructed using vendors' software from United Imaging to simulate the clinical reality. In the rest of this paper, instead of ``low-dose'', the term ``low-count'' will be used to be technically more accurate in a PET setting.

\section{Results}

\begin{table*}[!h]
\centering
\caption{Quantitative assessment for different methods. The measurements were obtained by averaging the values on the testing human studies. The proposed method consistently produced promising denoised results regardless of input count levels. $p<0.01$ was observed in all groups when comparing the proposed method with all other methods (except for 50\% SSIM values). The best results among different low-count levels are marked in \textcolor{red}{red}. Black voxels were removed for calculations.}
\resizebox{\textwidth}{!}{
\begin{tabular}{c|c|c|c|c|c|c}
\hline\hline
\multicolumn{7}{c}{\textbf{United Imaging uExplorer Scanner (20 patients $\times$ 6 = 120 studies)}}\\
\hline\hline
     \textbf{PSNR$\uparrow$/NRMSE$\downarrow$/SSIM$\uparrow$} &  1\% Count Input&   2\% Count Input&    5\% Count Input&  10\% Count Input& 25\% Count Input& 50\% Count  Input\\

\hline
Input  &44.667 / 0.682 / 0.788 & 49.260 / 0.404 / 0.895 & 53.575 / 0.249 / 0.957 & 56.085 / 0.189 / 0.977
 & 59.346 / 0.131 / 0.991 & 62.253 / 0.093 / 0.996 \\
\hline
UNN & 52.034 / 0.293 / 0.953 & 54.044 / 0.235 0.973 & 55.838 / 0.194 / 0.982 & 56.963 / 0.173 / 0.987
 & 58.570 / 0.146 / 0.991 &  59.885 / 0.126 / 0.994 \\
 \hline
DDIM50  & 42.550 / 0.852 / 0.904 & 42.718 / 0.836 / 0.926 & 42.846 / 0.824 / 0.938 & 42.942 / 0.815 / 0.944 & 43.065 / 0.804 / 0.952 & 43.145 / 0.797 / 0.957  \\
\hline
DDIM50+2.5D  & 43.636 / 0.754 0.910 & 43.851 / 0.736 / 0.924 & 43.856 / 0.735 / 0.930 & 43.829 / 0.738 / 0.932 &  43.805 / 0.740 / 0.936 & 43.820 / 0.738 / 0.938  \\
\hline
DiffusionMBIR  & 42.590 / 0.848 / 0.913 & 42.747 / 0.833 / 0.934 & 42.868 / 0.822 / 0.944 & 42.960 / 0.814 / 0.950 &  43.079 / 0.803 / 0.957 & 43.156 / 0.796 / 0.961  \\
\hline
DiffusionMBIR+2.5D  & 42.615 / 0.846 / 0.913 & 42.834 / 0.825 / 0.935 & 42.892 / 0.819 / 0.945 & 42.911 / 0.818 / 0.948 & 42.931 / 0.816 / 0.952 & 42.950 / 0.814 / 0.955  \\
\hline
TPDM  & 42.691 / 0.839 / 0.906 & 42.810 / 0.827 / 0.929 & 42.939 / 0.815 / 0.940 & 43.051 / 0.805 / 0.946 & 43.202 / 0.791 / 0.954 & 43.291 / 0.783 / 0.959 \\
\hline
TPDM+2.5D  & 42.334 / 0.873 / 0.887 & 42.458 / 0.861 / 0.912 & 42.483 / 0.859 / 0.923 & 42.499 / 0.857 / 0.927 & 42.528 / 0.854 / 0.932 & 42.536 / 0.853 / 0.935  \\
\hline
\hline
\textbf{DDPET-3D (proposed)}  & \textcolor{red}{52.899 / 0.267 / 0.965} & \textcolor{red}{54.937 / 0.215 / 0.977} & \textcolor{red}{57.119 / 0.171 / 0.985} & \textcolor{red}{58.551 / 0.148 / 0.989} & \textcolor{red}{60.916 / 0.117 / 0.993} & \textcolor{red}{63.804 / 0.088 / 0.996} \\
\hline
\end{tabular}
}
\label{table1}
\end{table*}

\begin{table*}[!h]
\centering
\caption{Quantitative assessment for different methods on all the testing patients from United Imaging scanner. The measurements were obtained by averaging the values on the testing human studies. The proposed method consistently produced promising denoised results regardless of input count levels. $p<0.01$ was observed in all groups when comparing the proposed method with all other methods (except for 50\% SSIM values). The best results among different low-count levels are marked in \textcolor{red}{red}. Black voxels were removed for calculations.}
\resizebox{\textwidth}{!}{
\begin{tabular}{c|c|c|c|c|c|c}
\hline\hline
\multicolumn{7}{c}{\textbf{United Imaging uExplorer Scanner (100 patients $\times$ 6 = 600 studies)}}\\
\hline\hline
     \textbf{PSNR$\uparrow$/NRMSE$\downarrow$/SSIM$\uparrow$} &  1\% Count Input&   2\% Count Input&    5\% Count Input&  10\% Count Input& 25\% Count Input& 50\% Count  Input\\

\hline
Input  & 43.141 / 0.706 / 0.780 & 47.727 / 0.418 / 0.889 & 52.052 / 0.254 / 0.954 & 54.535 / 0.191 / 0.975 & 57.481 / 0.135 / 0.990 & 59.805 / 0.103 / 0.995\\
\hline
UNN & 50.369 / 0.304 / 0.950 & 52.441 / 0.239 / 0.970 & 54.013 / 0.201 / 0.979 & 55.063 / 0.179 / 0.984 & 56.298 / 0.155 / 0.989 & 57.327 / 0.138 / 0.992 \\
 \hline
 \hline
\textbf{DDPET-3D (proposed)}  & \textcolor{red}{51.629 / 0.263 / 0.962} & \textcolor{red}{53.776 / 0.207 / 0.976} & \textcolor{red}{55.911 / 0.163 / 0.984} & \textcolor{red}{57.370 / 0.140 / 0.988} & \textcolor{red}{59.690 / 0.108 / 0.993} & \textcolor{red}{62.516 / 0.080 / 0.997} \\

\hline

\end{tabular}
}
\label{table2}
\end{table*}

\begin{figure*}[!t]
\centerline{\includegraphics[width=\textwidth]{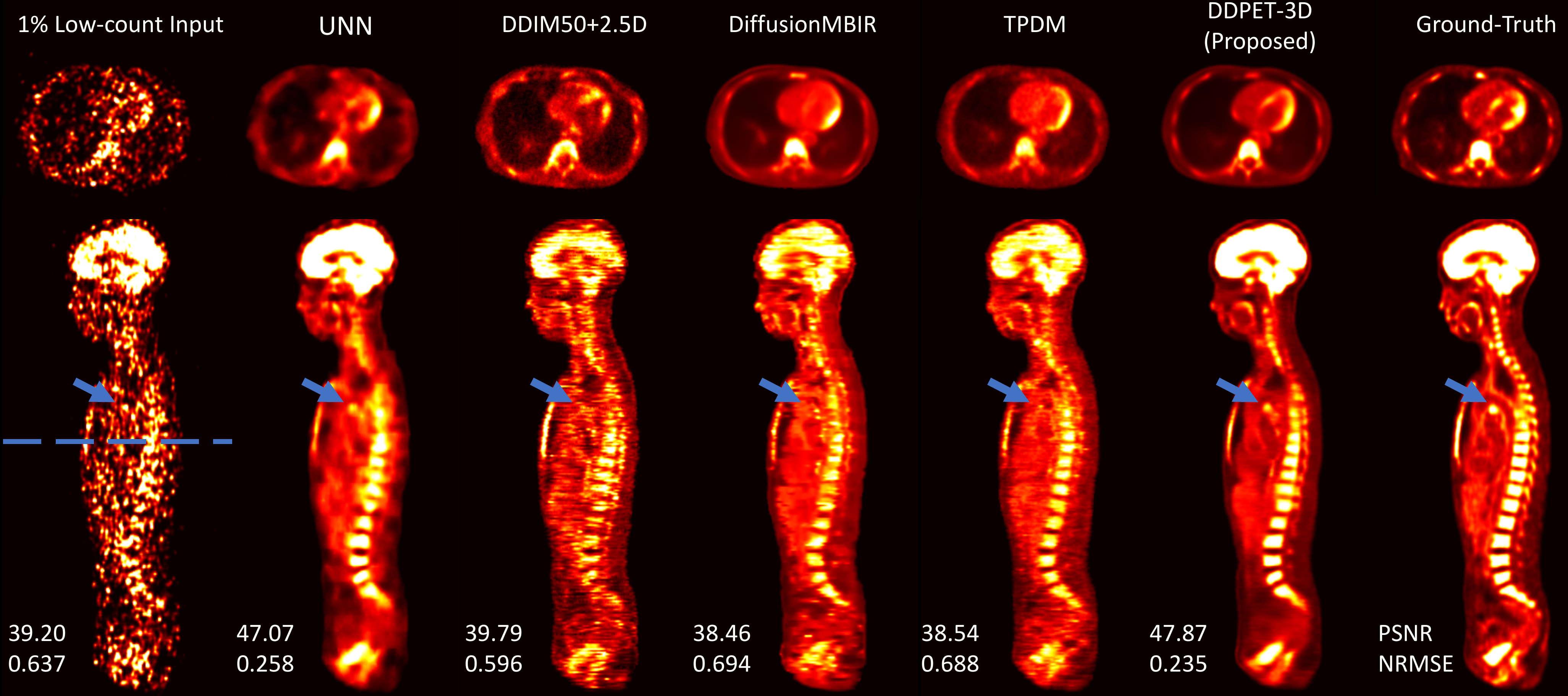}}
\caption{Low-count denoised images reconstructed using different methods. Note that all the images were normalized by the total injected activities. ``2.5D'' indicates multi-slice condition. Blue arrows point to a possible lesion that is better recovered by DDPET-3D. Compared to other baseline methods, DDPET-3D produced images with better details and maintain consistency along the slice dimension. Dashed blue line indicate the location of the transverse slice.}
\label{fig_results_2view}
\end{figure*}

\begin{figure*}[!t]
\centerline{\includegraphics[width=\textwidth]{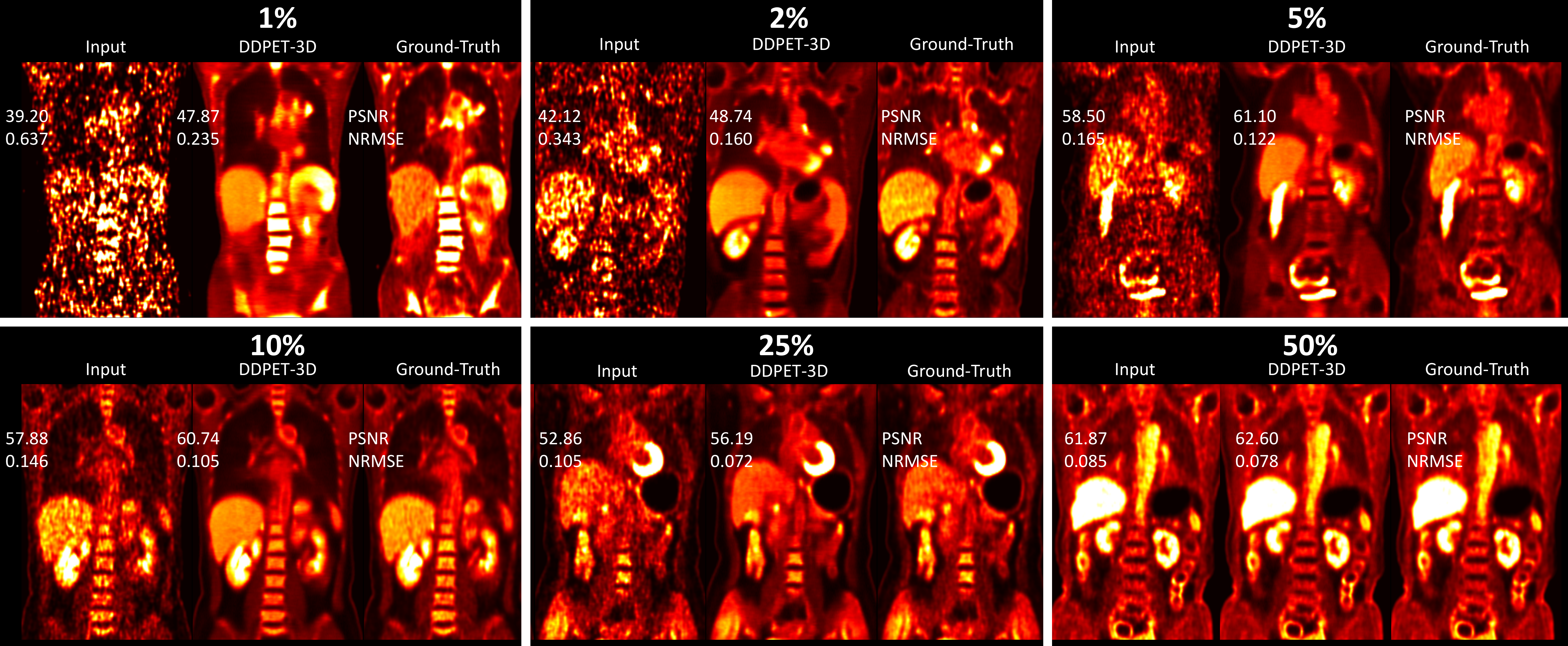}}
\caption{Low-count denoised images reconstructed using the proposed DDPET-3D with different low-count levels. 6 different patients at 6 different low-count levels are shown. DDPET-3D can produce consistent denoised results for a wide range of different noise-levels.}
\label{fig_results_all_doses_1view}
\end{figure*}

\begin{figure}[!t]
\centerline{\includegraphics[width=\linewidth]{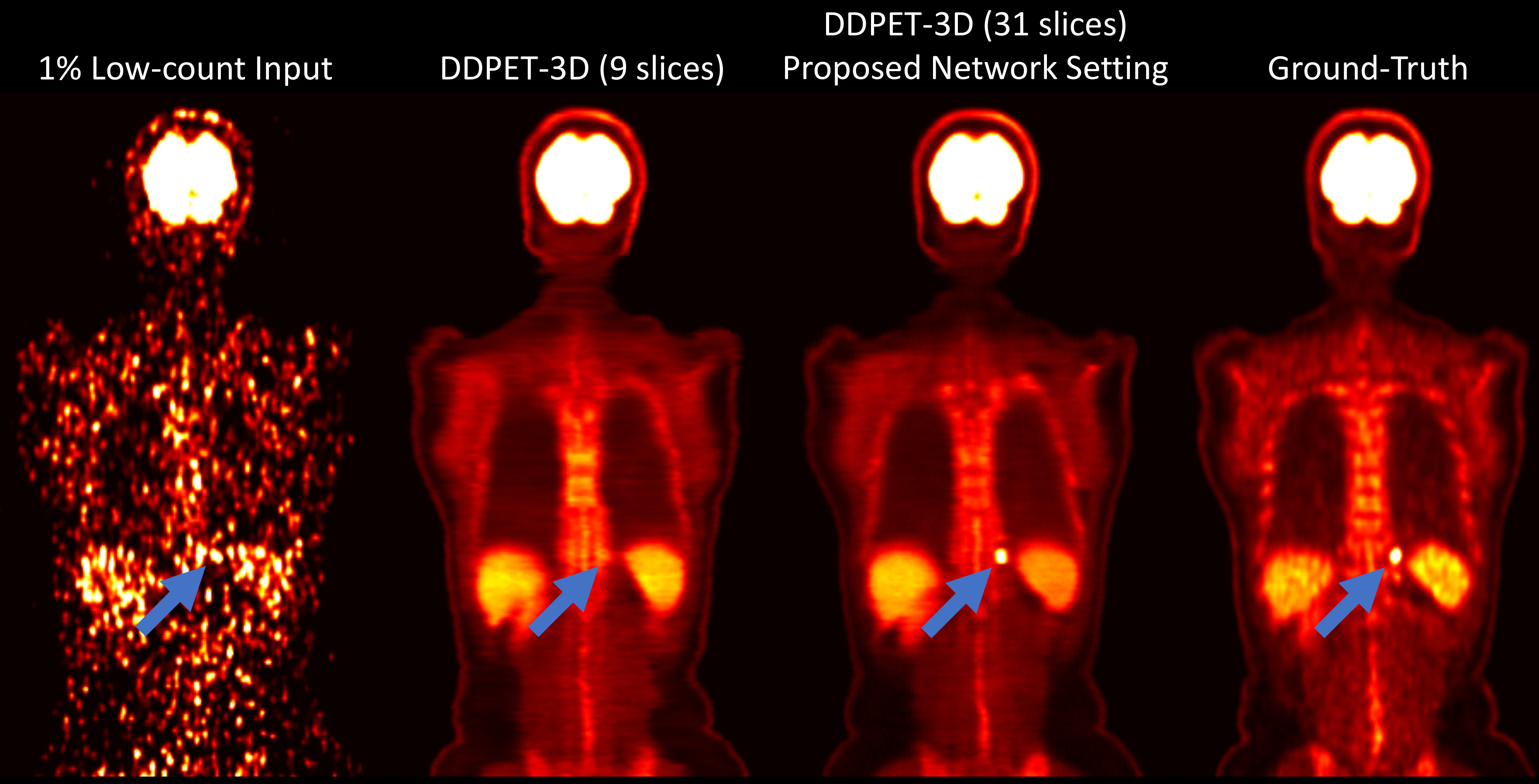}}
\caption{Images reconstructed using DDPET-3D with different numbers of conditional slices. Note that all the images were normalized by the total injected activities. Blue arrows point to a possible lesion that is better recovered by DDPET-3D with more conditional slices.}
\label{fig_results_num_slice}
\end{figure}

\begin{figure}[!t]
\centerline{\includegraphics[width=\linewidth]{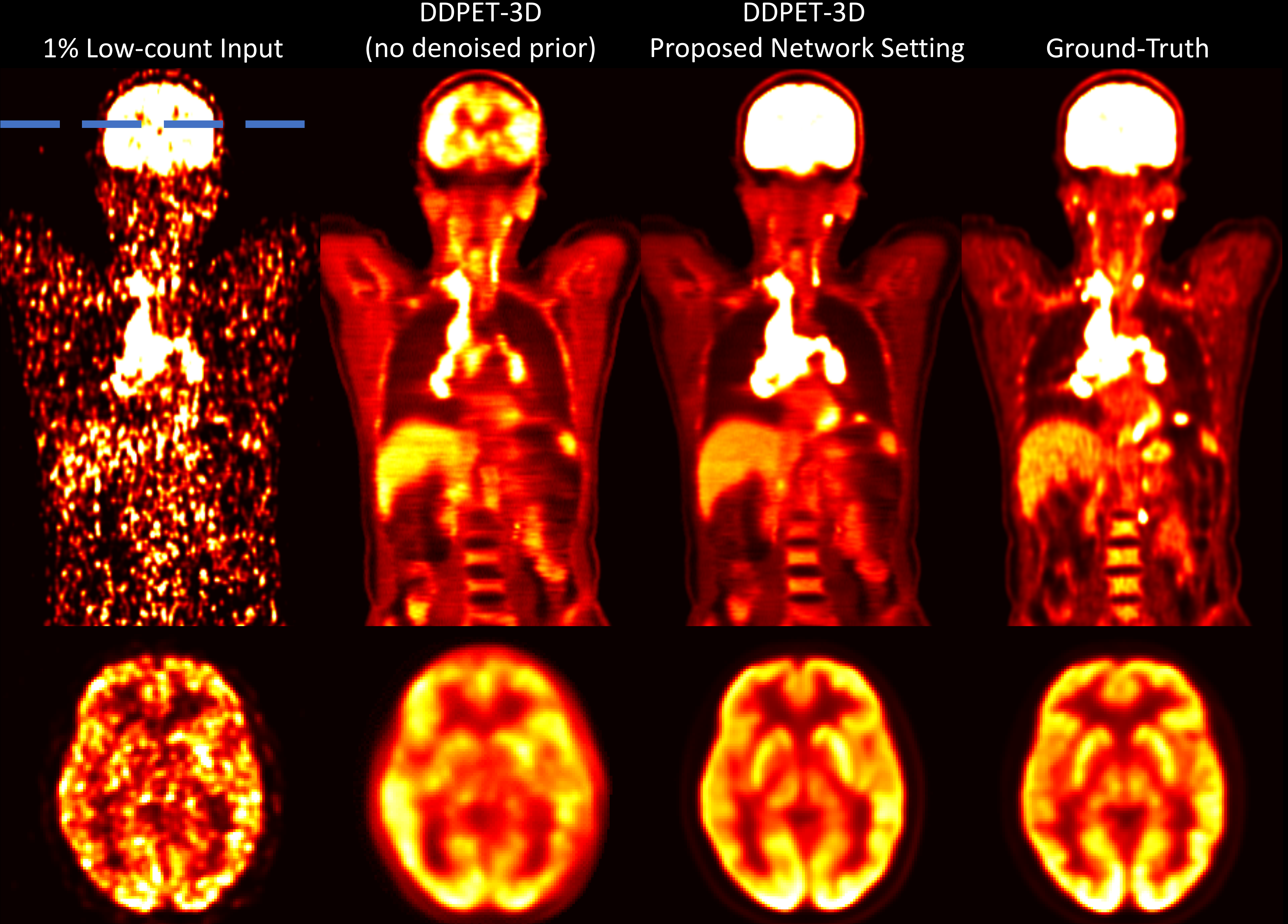}}
\caption{Visual comparison of images reconstructed using the proposed DDPET-3D methods with and without denoised priors. Note that all the images were normalized by the total injected activities. Note the tracer activity differences in different organs. For the brain transverse slice, due to quantification bias, image intensity in the "DDPET-3D (no prior)" method was manually tuned for visual comparison. Many image details were lost with no prior. Dashed blue line indicates the location of the transverse slice.}
\label{fig_results_no_unet}
\end{figure}

\begin{figure}[!t]
\centerline{\includegraphics[width=\linewidth]{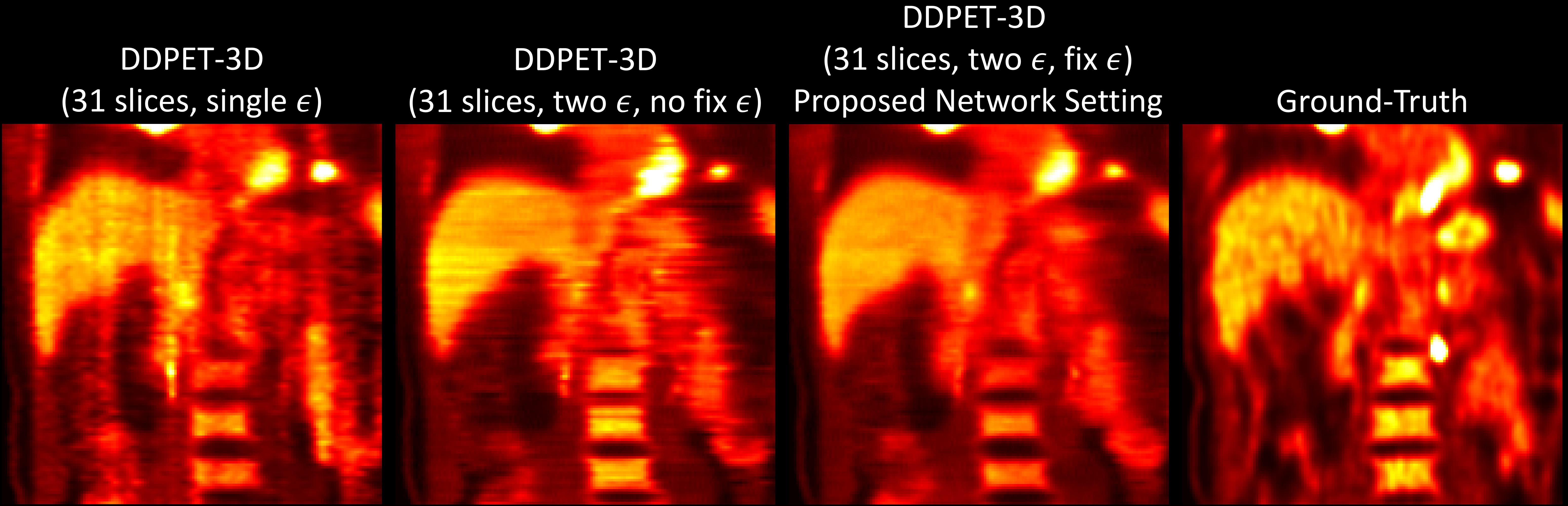}}
\caption{Images reconstructed with different settings of noise variables. Note that all the images were normalized by the total injected activities. Note the undesired artifacts along the z-direction by using only one noise variable $\epsilon$. Also note the inconsistencies of difference slices without fixing $\epsilon$. Despite similar quantitative values, proposed network settings produced images with more consistent 3D reconstructions and less artifacts.}
\label{fig_results_no_fix_seed_one_channel}
\end{figure}

% \subsection{Comparison with other methods}
\subsection{Baseline Comparison}
The proposed method was compared with DiffusionMBIR \cite{chung_solving_2023}, and TPDM \cite{lee_improving_2023}. Both of them were proposed to address the diffusion 3D inconsistencies problem in medical imaging. DiffusionMBIR applies a TV penalty term along the z-axis, and TPDM samples the images using 2 pre-trained perpendicular diffusion models. We also compared with the UNN method \cite{xie_unified_2023}, which was proposed for noise-aware PET denoising. Lastly, we compared with the standard DDIM sampling.

We extended the standard DDIM sampling method using conditional information from neighboring slices (denoted as DDIM50+2.5D). For sampling strategy, we used DDIM sampling for all diffusion models. Since our proposed method used 2 noise variables, for a fair comparison, all the comparison diffusion model used 50 sampling steps, while our proposed method only used 25 sampling steps (note that the proposed method runs 26 steps with 2 noise variables, since it only averages the 2 outputs from the first reverse step). 

As presented in Fig.~\ref{fig_results_2view}, even though we used neighboring slices as conditional information for DDIM sampling, both DiffusionMBIR and TPDM still produced more consistent reconstructions along the z-axis. However, the inconsistent issue still exist in both images. Proposed DDPET-3D method produced noticeable more consistent reconstructions by simply fixing the noise variables in the reverse process. Moreover, in the transverse plane, despite the promising visual results, all other comparison diffusion models produced images with distorted organs (e.g., myocardium in the transverse slice), which is unacceptable in clinical settings. The U-net-based method (UNN), on the other hand, still maintained the overall structure of the heart.

Another major issue of diffusion model is inaccurate image quantification. Note that all the presented images were already normalized by the total injected activities. As shown in Fig.~\ref{fig_results_2view}, the tracer activities in certain organs are completely wrong in other comparison diffusion models. For example, the activities in the brain are noticeably lower in DDIM, DiffusionMBIR, and TPDM results. Such difference may affect certain diagnostic tasks such as lesion detection \cite{schaefferkoetter2015initial}. Even though UNN produced over-smoothed reconstructions, it does not alter the overall tracer activities in different organs. Using UNN output as denoised prior, the proposed DDPET-3D maintained overall image quantification. Also, using UNN as a denoised prior allows the proposed method to recover some subtle features that are almost invisible in the low-count input (blue arrows in Fig.~\ref{fig_results_2view}).

Images were quantitatively evaluated using SSIM (Structural Similarity Index), PSNR (Peak signal-to-noise ratio), and RMSE (Root-mean-square error). To facilitate the testing process, we tested all the comparison methods using 20 patients from the entire testing dataset. There are 6 different low-count levels of all the patients, resulting in a total of 120 testing studies. Quantitative results are presented in Table \ref{table1}. We also extended both DiffusionMBIR and TPDM, and re-trained them using neighboring slices as conditional information (denoted as DiffusionMBIR+2.5D and TPDM+2.5D in Table \ref{table1}).

As presented in Table \ref{table1}, the proposed method outperformed other methods in all the 6 low-count levels. When the input count-level increases, the performance of all the methods gradually improves. Both the proposed method and UNN can achieve noise-aware denoising. However, at higher count-levels (25\% and 50\%), UNN results were even worse than the input. In contrast, the proposed method produced optimal results at all count-levels. Also, due to inaccurate image quantification, other diffusion models produced images with even worse quantitative results compared with low-count inputs at higher count-levels.

It is worth noting that, simply adding neighboring slices as conditional information does not necessarily lead to better performance for DiffusionMBIR and TPDM. Visual comparisons for DiffusionMBIR+2.5D, TPDM+2.5D, and other methods are presented in Supplemental Fig.~\ref{supp_fig_results}.

As presented in Table \ref{table1}, UNN is the second-best method. To comprehensively evaluate the proposed method, We compared the proposed method with UNN for all the 100 testing patients. As presented in Table \ref{table2}, DDPET-3D consistently outperformed UNN on different low-count levels.

To demonstrate that DDPET-3D achieves dose-aware denoising, denoised results using inputs with different low-count levels are presented in Fig. \ref{fig_results_all_doses_1view}. DDPET-3D consistently produced promising denoised results regardless of input count-levels. The reconstruction results gradually improve when the input count-level increases. Some subtle structures are better visualized in denoised results with higher count-levels input images.

\subsection{Ablation Studies}

We performed several ablated experiments to demonstrate the effectiveness of different proposed components in DDPET-3D.

\textbf{Impact of Number of Conditional Slices:} We evaluated the DDPET-3D with different numbers of neighboring conditional slices. Three variants of DDPET-3D networks were trained using 9, 21, and 41 neighboring slices as conditional information. These networks are denoted as "DDPET-3D ($n$ slice)", where $n$ is the number of conditioned neighboring slices. Results showed that $n=31$ performed the best in most quantitative metrics. Therefore, $n=31$ was used in this paper. Quantitative results are presented in Supplemental Table \ref{table3}. As presented in Fig.~\ref{fig_results_num_slice}, using more conditioned slices helped the network to recover some subtle details in the images (blue arrows in Fig.~\ref{fig_results_num_slice}).

% In the first experiment, we tested the DDPET-3D method with different number of neighboring conditioned slices. Three variants of DDPET-3D networks were trained using 9, 21, and 41 neighboring slices as conditional information. These networks are denoted as "DDPET-3D ($n$ slice)", where $n$ is the number of conditioned neighboring slices. Results showed that $n=31$ performed the best in most quantitative metrics. Therefore, $n=31$ was used in this paper. Quantitative results are presented in Supplemental Table \ref{table3}. As presented in Fig.~\ref{fig_results_num_slice}, using more conditioned slices helped the network to recover some subtle details in the images (blue arrows in Fig.~\ref{fig_results_num_slice}).

\textbf{Impact of Denoised Prior:} We tested the DDPET-3D without the denoised prior during the sampling process to demonstrate the improvement in image quantification using this denoised prior. This network is denoted as "DDPET-3D (no prior)" in Supplemental Table \ref{table3}. As presented in Fig.~\ref{fig_results_no_unet}, we can see that diffusion models produced images with inaccurate tracer activities in different organs without the proposed denoised prior (especially in the brain and liver). Note that the images were already normalized by the total injected activities of the entire 3D volume. Also, without the denoised prior, many details in the images were not able to recovered, as presented in the brain slice in Fig.~\ref{fig_results_no_unet}. 

% In the second experiment, we tested the DDPET-3D method without the denoised prior during the sampling process to demonstrate the improvement in image quantification using this denoised prior. This network is denoted as "DDPET-3D (no prior)" in Supplemental Table \ref{table3}. As presented in Fig.~\ref{fig_results_no_unet}, we can see that diffusion models produced images with inaccurate tracer activities in different organs without the proposed denoised prior (especially in the brain and liver). Note that the images were already normalized by the total injected activities of the entire 3D volume. Also, without the denoised prior, many details in the images were not able to recovered, as presented in Fig.~\ref{fig_results_no_unet_brain}. 

\textbf{Impact of Fixing Noise Variables:} We analyzed the DDPET-3D with and without fixing the 2 noise variables to demonstrate its effectiveness in producing consistent 3D reconstructions. Specifically, $\epsilon_0^a$ and $\epsilon_0^b$ are sampled from $\Ncal(0,\textbf{I})$ for every slice in the 3D image volume, instead of fixing them for all slices. This network is denoted as "DDPET-3D (no fix $\epsilon$)" in Supplemental Table \ref{table3}. Although the differences in quantitative evaluations with and without fixing noise variables are small, and not fixing noise even led to better quantitative measurements in certain cases, we noticed significant improvements in visual quality, as presented in Fig.~\ref{fig_results_no_fix_seed_one_channel}. Not fixing noise variables produced images with inconsistent slices and unclear organ boundaries, which are unfavorable in clinical settings.

% In the third experiment, we tested the DDPET-3D method without fixing the 2 noise variables to demonstrate the effectiveness of it to produce consistent 3D reconstructions. Specifically, $\epsilon_0^a$ and $\epsilon_0^b$ are sampled from $\Ncal(0,\textbf{I})$ for every slice in the 3D image volume, instead of fixing them for all slices. This network is denoted as "DDPET-3D (no fix $\epsilon$)" in Supplemental Table \ref{table3}. Although the differences in quantitative evaluations with and without fixing noise variables are small, and not fixing noise even led to better quantitative measurements in certain cases, we noticed significant improvements in visual quality, as presented in Fig.~\ref{fig_results_no_fix_seed_one_channel}. Not fixing noise variables produced images with inconsistent slices and unclear organ boundaries, which are unfavorable in clinical settings.

\textbf{Impact of Using Multiple Noise Variables:} We tested the DDPET-3D method with only one noise variable. Specifically, we only initialized one noise variable instead of 2 ($\epsilon_0^a$ and $\epsilon_0^b$). This method is denoted as "DDPET-3D (single $\epsilon$)" in Supplemental Table \ref{table3}. Using 2 noise variables produced images with better quantitative assessments. Also, Fig.~\ref{fig_results_no_fix_seed_one_channel} shows that using only one noise variable would result in undesired artifacts along the z-direction.

% In the fourth experiment, we tested the DDPET-3D method with only one noise variable. Specifically, we only initialize one noise variable instead of 2 ($\epsilon_0^a$ and $\epsilon_0^b$). This method is denoted as "DDPET-3D (single $\epsilon$)" in Supplemental Table \ref{table3}. Using 2 noise variables produced images with better quantitative assessments. Also, as presented in Fig.~\ref{fig_results_no_fix_seed_one_channel}, using only one noise variable would result in some undesired artifacts along the z-direction.

\textbf{Impact of Dose Embedding:} We tested the DDPET-3D without dose embedding, which is denoted as "DDPET-3D (no dose)" in Supplemental Table \ref{table3}. Experimental results showed that DDPET-3D with dose embedding produced images with better quantitative results. 

% Lastly, we tested the DDPET-3D method without dose embedding. This method is denoted as "DDPET-3D (no dose)" in Supplemental Table \ref{table3}. Experimental results showed that DDPET-3D with dose embedding produced images with better quantitative results. 

\textbf{Generalizability Test:} To evaluate DDPET-3D's generalizability, we applied the trained model to patient data acquired using a different scanner from a different hospital in another country. 20 patient studies were acquired using the Siemens Vision Quadra scanner at XX in this experiment. Results are shown in Supplemental Fig. \ref{fig_results_quadra_using_UI_model} and Supplemental Table \ref{table4}.
\section{Discussion and Conclusion}
We proposed DDPET-3D, a dose-aware diffusion model for 3D ultra-low-dose PET imaging. Compared with previous methods, DDPET-3D has the following contributions: \emph{(1)} 3D reconstructions with fine details and address the 3D inconsistency issue in previous diffusion models. This was achieved by using multiple neighboring slices as conditional information and enforcing the same Gaussian latent for all slices in sampling. \emph{(2)} DDPET-3D maintains accurate image quantification by using a pre-trained denoised prior in sampling. \emph{(3)} DDPET-3D achieves dose-aware denoising. It can be generalized to different low-count levels, ranging from 1\% to 50\%. \emph{(4)} With all the proposed strategies, DDPET-3D converges within 25 diffusion steps, allowing fast reconstructions. It takes roughly 15 minutes to reconstruct the entire 3D volume, while DDPM sampling takes about 6 hours. The experimental results showed that DDPET-3D achieved SOTA performance compared with previous methods. In addition, for the first time, we showed that diffusion model can gain promising results on 1\% ultra-low-dose PET denoising problem, both quantitatively and qualitatively. While this study focuses on PET image denoising, we believe that the proposed method could be easily extended to other 3D reconstruction tasks for different imaging modalities.

{
    \small
    \bibliographystyle{ieeenat_fullname}
    \bibliography{main}
}
 \clearpage
\setcounter{page}{1}
\maketitlesupplementary

\begin{figure*}[t]
\centering
{\includegraphics[width=\textwidth]{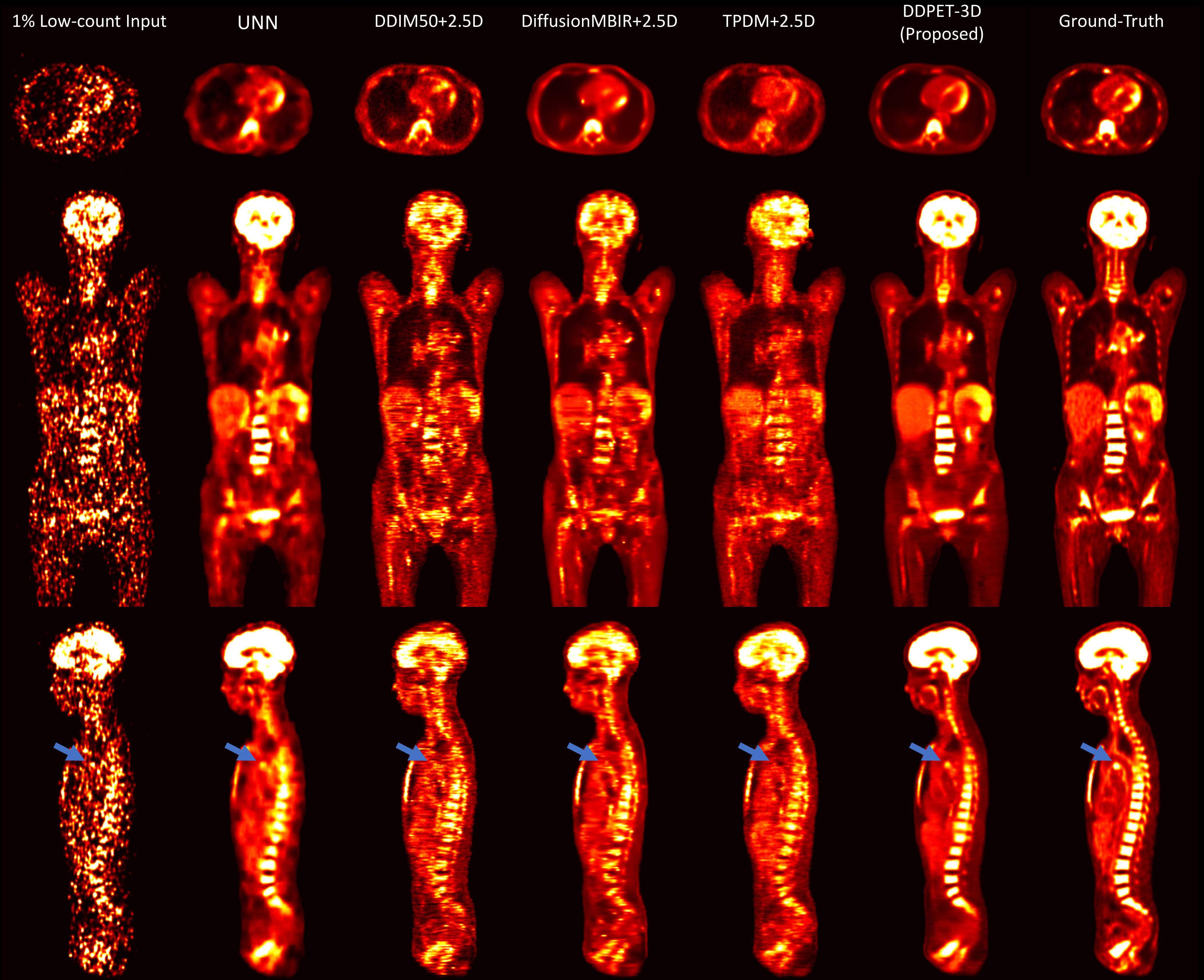}}
\caption{Visual comparison of low-count denoised images reconstructed using different methods. Note that all the images were normalized by the total injected activities. "2.5D" indicates multi-slice condition.}
\label{supp_fig_results}
\end{figure*}

Fig. \ref{supp_fig_results}: we extended DiffusionMBIR and TPDM methods using neighboring slices as conditional information (denoted as DiffusionMBIR+2.5D and TPDM+2.5D). 

\begin{figure*}[!t]
\centerline{\includegraphics[width=\textwidth]{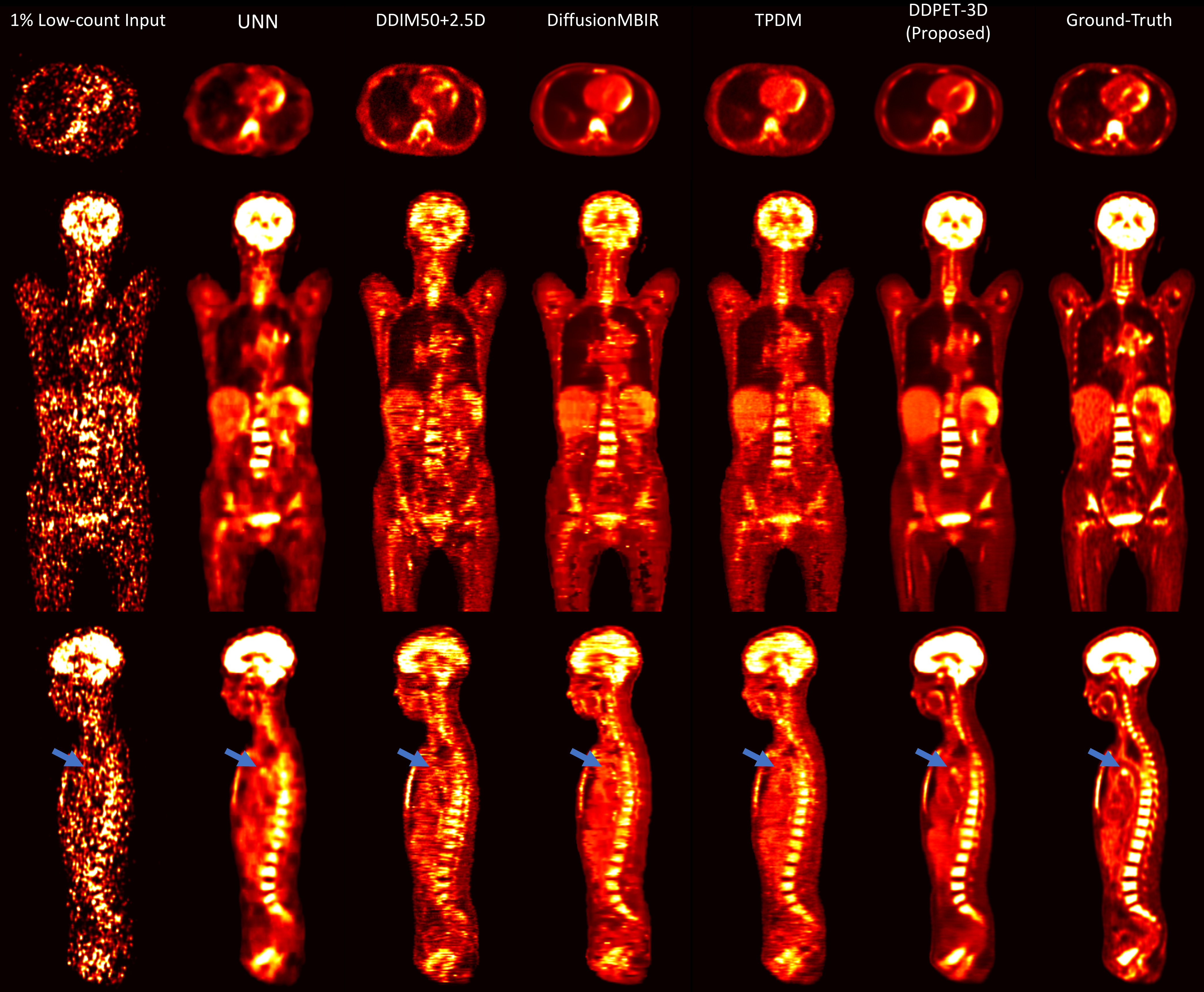}}
\caption{Visual comparison of low-count denoised images reconstructed using different methods. Note that all the images were normalized by the total injected activities. ``2.5D'' indicates multi-slice condition. Blue arrows point to subtle details that are better recovered by the proposed method. It is worth highlighting that, compared to other baseline methods, DDPET can reconstruct images with fine-details and maintain consistency along the slice dimension.}
\label{fig_results}
\end{figure*}

Fig. \ref{fig_results}: we added coronal view in addition to Fig. \ref{fig_results_2view} in the main text.

\begin{figure*}[!t]
\centerline{\includegraphics[width=\textwidth]{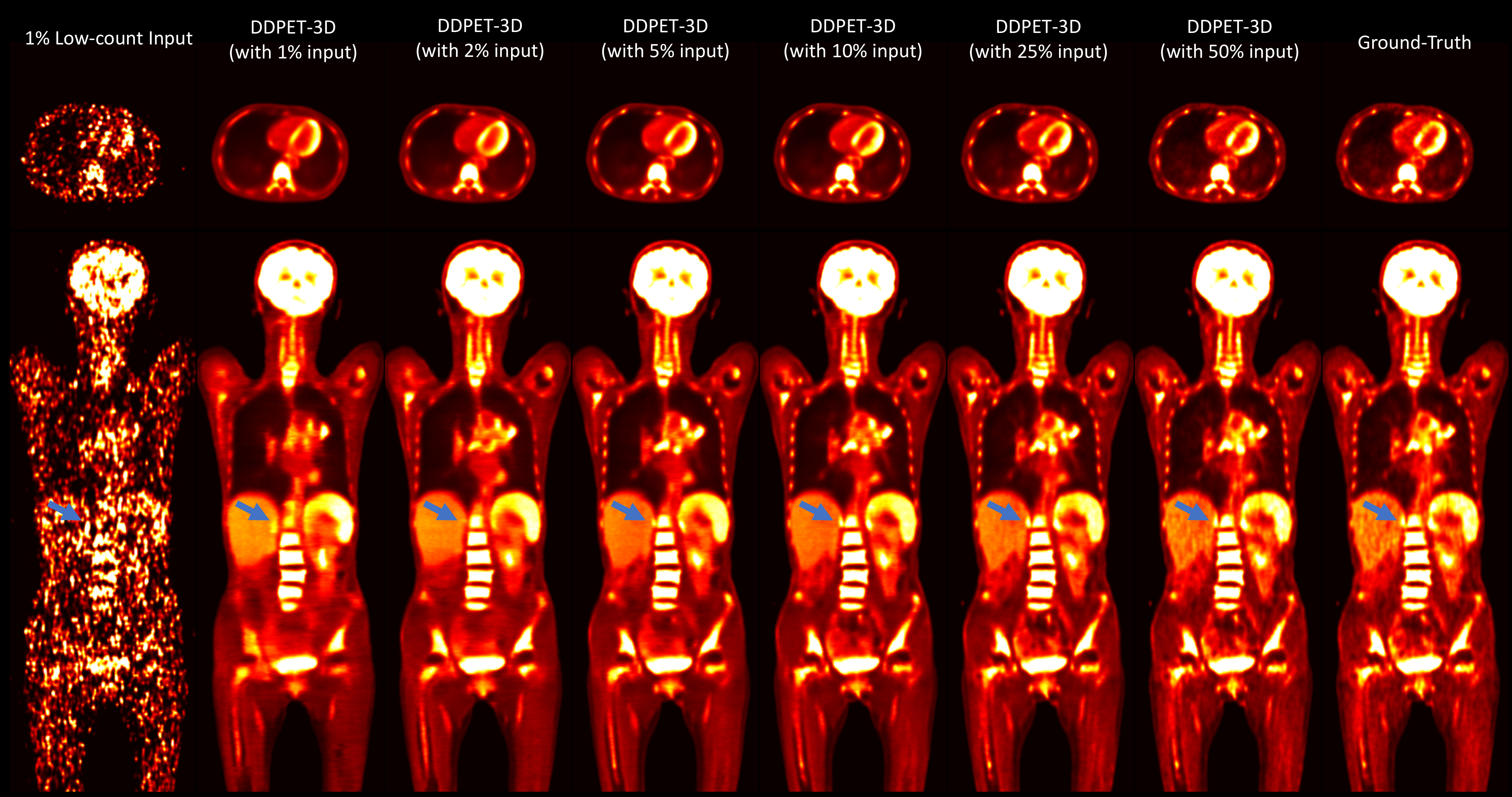}}
\caption{Visual comparison of low-count denoised images reconstructed using the proposed DDPET-3D method with different low-count levels. Blue arrows point to subtle details that are gradually better recovered with higher-count inputs. DDPET-3D can produce consistent denoised results for a wide range of different noise-levels.}
\label{fig_results_all_doses}
\end{figure*}

Fig. \ref{fig_results_all_doses}: Images reconstructed using DDPET-3D with inputs with different low-count levels.

\begin{figure*}[!t]
\centerline{\includegraphics[width=\textwidth]{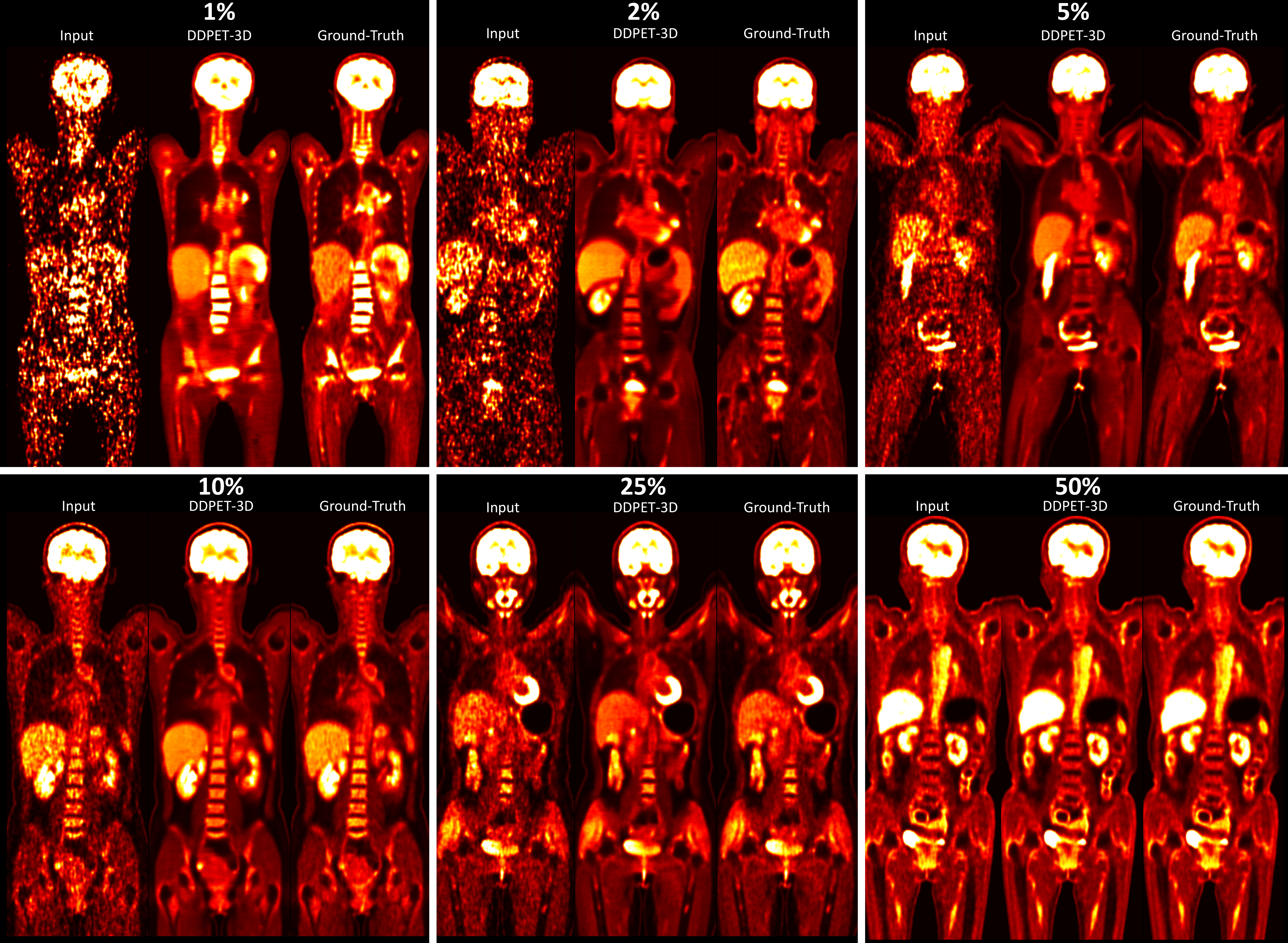}}
\caption{Low-count denoised images reconstructed using the proposed DDPET-3D method with different low-count levels. DDPET-3D can produce consistent denoised results for a wide range of different noise-levels.}
\label{fig_results_all_doses_1view_no_crop}
\end{figure*}

Fig. \ref{fig_results_all_doses_1view_no_crop}: Same patients presented in Fig. \ref{fig_results_all_doses_1view} with larger field-of-view.

\begin{figure*}[!t]
\centerline{\includegraphics[width=0.6\textwidth]{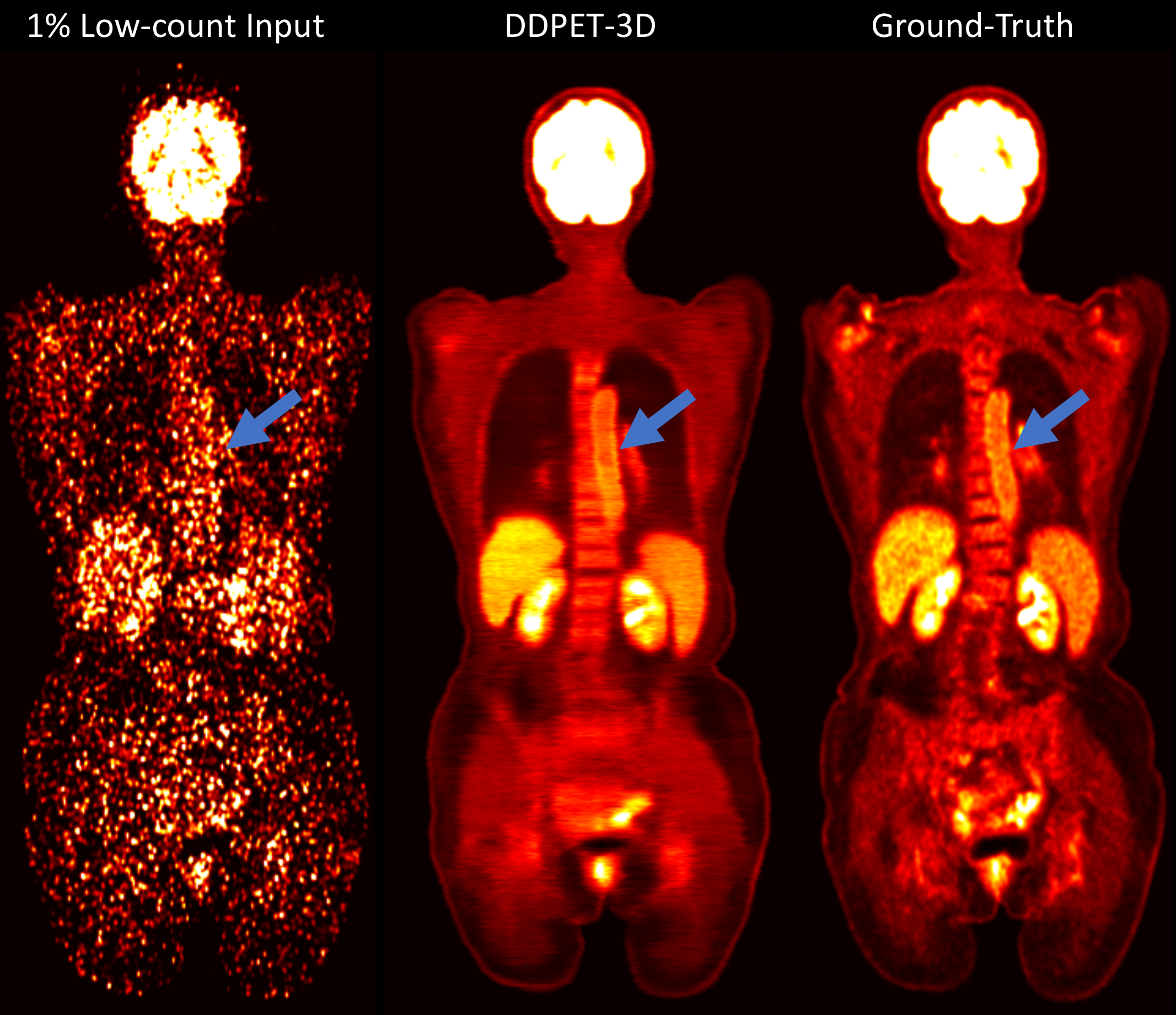}}
\caption{Patient data acquired on a Siemens Vision Quadra scanner at Country YY reconstructed using the proposed DDPET-3D. Specifically, the DDPET-3D was trained using data acquired on the United Imaging uExplorer scanner at Country XX. The presented results demonstrate the strong generalizability of the proposed method. As pointed out by the blue arrow, the aorta can be recovered from the 1\% low-count input.}
\label{fig_results_quadra_using_UI_model}
\end{figure*}

Fig. \ref{fig_results_quadra_using_UI_model}: to demonstrate that the proposed DDPET-3D method has superior generalizability, we directly apply the trained model on patient data acquired on a different scanner from another hospital. 20 patient studies were used for this experiment.

Table \ref{table3}: quantitative assessments of different ablation studies described in the main text.

Table \ref{table4}: quantitative assessments of the generalizability test described in the main text.

\begin{table*}[!h]
\centering
\caption{Quantitative assessment for different ablation studies. The best results among different low-count levels are marked in \textcolor{red}{red}. The second best results are marked in \textcolor{blue}{blue}. Black voxels were removed for calculations.}
\resizebox{\textwidth}{!}{
\begin{tabular}{c|c|c|c|c|c|c}
\hline\hline
\multicolumn{7}{c}{\textbf{United Imaging uExplorer Scanner (20 patients $\times$ 6 = 120 studies)}}\\
\hline\hline
     \textbf{PSNR$\uparrow$/NRMSE$\downarrow$/SSIM$\uparrow$} &  1\% Count Input&   2\% Count Input&    5\% Count Input&  10\% Count Input& 25\% Count Input& 50\% Count  Input\\

\hline
DDPET-3D (9 slices) & 52.563 / 0.277 / 0.964 & 54.666 / 0.215 / \textcolor{red}{0.977} & 56.194 / \textcolor{red}{0.162} / 0.984 & 57.688 / \textcolor{red}{0.137} / 0.988 & 60.011 / \textcolor{red}{0.106} / \textcolor{red}{0.993} & 62.267 / \textcolor{blue}{0.082} / 0.995 \\
 \hline
DDPET-3D (21 slices)  & 52.502 / 0.278 / 0.960 & 54.509 / 0.224 / 0.974 & 56.574 / 0.181 / 0.981 & 57.949 / 0.157 / 0.985 & 60.137 / 0.125 / 0.990 &  62.338 / 0.100 / 0.994 \\
\hline
DDPET-3D (41 slices)  & 51.866 / \textcolor{red}{0.262} / 0.964 & 53.926 / \textcolor{red}{0.208} / \textcolor{red}{0.977} & 56.147 / \textcolor{red}{0.162} / \textcolor{red}{0.985} & 57.618 / \textcolor{blue}{0.138} / \textcolor{red}{0.989} & 59.948 / \textcolor{blue}{0.107} / \textcolor{red}{0.993} &  62.438 / \textcolor{red}{0.081} / 0.995 \\
\hline
DDPET-3D (no prior)  & 41.497 / 0.961 / 0.868 & 41.638 / 0.945 / 0.898 & 41.704 / 0.938 / 0.913 & 41.715 / 0.937 / 0.920 & 41.747 / 0.934 / 0.927 & 41.767 / 0.932 / 0.932 \\
\hline
DDPET-3D (no fix $\epsilon$)  & \textcolor{red}{52.991} / \textcolor{blue}{0.264} / 0.961 & \textcolor{red}{55.049} / \textcolor{blue}{0.212} / 0.973 & \textcolor{red}{57.225} / \textcolor{blue}{0.170} / 0.980 & \textcolor{red}{58.695} / 0.146 / 0.983 & \textcolor{red}{61.026} / 0.116 / 0.988 &  \textcolor{red}{63.916} / 0.087 / 0.991 \\
\hline
DDPET-3D (no dose)  & 52.817 / 0.269 / 0.960 & 54.822 / 0.217 / 0.972 & 56.903 / 0.175 / 0.979 & 58.259 / 0.152 / 0.983 & 60.360 / 0.123 / 0.987 &  62.871 / 0.096 / 0.990 \\
\hline
DDPET-3D (single $\epsilon$)  & 52.289 / 0.285 / 0.959 & 54.261 / 0.231 / 0.974 & 56.324 / 0.185 / 0.983 & 57.701 / 0.161 / 0.987 & 59.917 / 0.128 / 0.992 & 62.548 / 0.098 / 0.995 \\
\hline
\textbf{DDPET-3D (proposed)}  & \textcolor{blue}{52.899} / 0.267 / \textcolor{red}{0.965} & \textcolor{blue}{54.937} / 0.215 / \textcolor{red}{0.977} & \textcolor{blue}{57.119} / 0.171 / \textcolor{red}{0.985} & \textcolor{blue}{58.551} / 0.148 / \textcolor{red}{0.989} & \textcolor{blue}{60.916} / 0.117 / \textcolor{red}{0.993} & \textcolor{blue}{63.804} / 0.088 / \textcolor{red}{0.996} \\
\hline
\end{tabular}
}
\label{table3}
\end{table*}

\begin{table*}[!h]
\centering
\caption{Quantitative assessment for the generalizability test. The model was trained using the United Imaging uExplorere scanner at Country XX and directly applied to 20 patient studies acquired using a Siemens Vision Quadra scanner at Country YY. The best results among different low-count levels are marked in \textcolor{red}{red}. 50\% count images were not available for these patient studies. Black voxels were removed for calculations.}
\resizebox{\textwidth}{!}{
\begin{tabular}{c|c|c|c|c|c}
\hline\hline
\multicolumn{6}{c}{\textbf{Siemens Vision Quadra Scanner (20 patients $\times$ 5 = 100 studies)}}\\
\hline\hline
     \textbf{PSNR$\uparrow$/NRMSE$\downarrow$/SSIM$\uparrow$} &  1\% Count Input&   2\% Count Input&    5\% Count Input&  10\% Count Input& 25\% Count Input\\

\hline
Input  & 50.964 / 0.385 / 0.904 & 53.737 / 0.275 / 0.946 & 56.993 / 0.188 / 0.976 & 59.273 / 0.145 / 0.987 & 62.708 / 0.098 / 0.995  \\
 \hline
\textbf{DDPET-3D (proposed)}  & \textcolor{red}{56.818 / 0.192 / 0.981} & \textcolor{red}{58.130 / 0.165 / 0.986} & \textcolor{red}{59.604 / 0.139 / 0.991}  & \textcolor{red}{60.692 / 0.123 / 0.993} & \textcolor{red}{62.850 / 0.096 / 0.996} \\
\hline
\end{tabular}
}
\label{table4}
\end{table*}

\end{document}